\documentclass[12pt]{article}
\usepackage{amsmath, amssymb, amsfonts, amsthm}
\usepackage{mathtools}
\usepackage{fullpage, setspace}
\usepackage{bm}
\usepackage{booktabs}
\usepackage{natbib}
\usepackage{color}
\usepackage{graphicx} 
\usepackage{changepage}
\usepackage{enumitem} 
\usepackage[affil-it]{authblk}

\newtheorem{proposition}{Proposition}
\newtheorem{definition}{Definition}
\newtheorem{remark}{Remark}

\title{Informed Bayesian Finite Mixture Models via Asymmetric Dirichlet Priors}

\author[1]{Garritt L. Page}
\author[2]{Massimo Ventrucci}
\author[3]{Maria Franco-Villoria}
\affil[1]{Department of Statistics, Brigham Young University, Provo, USA}
\affil[2]{Department of Statistical Sciences, University of Bologna, Bologna, Italy}
\affil[3]{Department of Economics “Marco Biagi”, University of Modena and Reggio Emilia, Italy}
\date{}

\begin{document}

\maketitle
\begin{abstract}
Finite mixture models are flexible methods that are commonly used for model-based clustering. A recent focus in the model-based clustering literature is to highlight the difference between the number of components in a mixture model and the number of clusters. The number of clusters is more relevant from a practical stand point, but to date, the focus of prior distribution formulation has been on the number of components.  In light of this, we develop a finite mixture methodology that permits eliciting prior information directly on the number of clusters in an intuitive way.  This is done by employing an asymmetric Dirichlet distribution as a prior on the weights of a finite mixture.  Further, a penalized complexity motivated prior is employed for the Dirichlet shape parameter.  We illustrate the ease to which prior information can be elicited via our construction and the flexibility of the resulting induced prior on the number of clusters.  We also demonstrate the utility of our approach using numerical experiments and two real world data sets. 
\end{abstract}

\noindent%
{\it Keywords:}  Bayesian clustering, Penalized Complexity Priors, Functional Data, Number of clusters 

\doublespacing

\section{Introduction}
Finite mixture models (FMMs) have become a popular tool in, among other things,  density estimation and unsupervised learning (i.e.,  model-based clustering).  An underlying assumption of FMMs is that each unit's measured realization comes from one of $K$ subgroups with group membership unknown {\it a priori}.    The realizations from each of the $K$ subgroups are then modeled with an appropriate density.  This produces a procedure that is able to accommodate distributions that cannot be modeled satisfactorily with a parametric model.  In its most general form a FMM can be expressed as
\begin{align} \label{fmm}
f(\bm{y}_i | \bm{\theta}_1, \ldots, \bm{\theta}_K, \bm{w}) = \sum_{k=1}^K w_k f_k(\bm{y}_i|\bm{\theta}_k),
\end{align}
where $\bm{w} = (w_i , \ldots, w_K)$  are component weights such that $\sum_{k=1}^K w_k = 1$,  $\bm{\theta} = (\bm{\theta}_1, \ldots, \bm{\theta}_K)$ component specific parameters, and $f_k(\cdot)$ a well defined component density.  From a Bayesian perspective, the model is finished by assigning prior distributions to $\bm{w}$, $\bm{\theta}$ and possibly $K$. A key reason why FMMs like those in \eqref{fmm} have garnered attention is due to their extreme flexibility with regards to the shape of $f$ which seamlessly permits their use in a diverse array of applications. However, there is a cost to this flexibility as the clustering arising from FMMs can be quite delicate to model specifications (e.g., prior distributions for $\bm{\theta}$, $\bm{w}$, or $K$).   Since prior decisions can have a significant impact on clustering  and common noninformative priors are known to perform poorly, it would be very appealing to construct a method that connects scientifically relevant prior information to meaningful model quantities. As a result, users would be able to more easily inform and regulate the FMM.

Decisions about $K$ are particularly impactful to a FMM's model fit.  As such, significant attention has been dedicated to studying it.  In the Bayesian FMM literature two approaches have emerged.  One is to treat $K$ as an unknown, random quantity, to which a prior distribution is assigned.   \cite{miller:2018} have referred to this approach as a mixture of finite mixture models (MFMM).  Until recently, most attempts to employ a MFMM required constructing a customized reversible-jump MCMC algorithm (RJMCMC; \citealt{Richardson:1997}) which called for a high level of expertise.  Because of this, the approach in \cite{Richardson:1997} was unavailable to many practitioners (for an alternative to RJMCMC see \citealt{stephens:2000}).  Recently, \cite{miller:2018} connected MFMMs to random partition models.  This made available the computational techniques developed in the Bayesian nonparametric (BNP) literature making MFMMs more accessible.  

The second approach prescribes formulating an overparametrized FMM by setting $K$ to a  large value and using a prior on $\bm{w}$ that ``shrinks'' some of the component weights to zero. \cite{RousseauMengersen:2011} have provided some theoretical justification for this approach which \cite{walli:2016} refer to as a sparse finite mixture model (sFMM). An alternative approach of producing a sFMM is to use a repulsive type prior on the parameter of centrality in $\bm{\theta}$.  See \cite{NIPS2012_4589}, \cite{xie&xu:20}, \cite{quinlan:2021}, \cite{Beraha_etal:2022}, and \cite{sun_etal:2022}.  

When $K$ is fixed at a large value and empty components are expected, it is straightforward to distinguish between the number of mixture components and the number of (data informed) clusters (which we denote as $K^+$).   However, with $K$ unknown, it is less clear and until recently it was generally thought that $K=K^+$.  There is now an emerging FMM literature that explicitly addresses the difference between $K$ and $K^+$ (\citealt{schnatter:2019, greve:2020, schnatter_etal:2021, quinlan:2021, argiento:2022, alamichel2023bayesian}).  As a consequence, the prior distribution of $K^+$ induced by particular FMM modeling decisions has begun to garner attention. 

\begin{figure}
\centering
\includegraphics[width=0.45\textwidth, trim={0 0cm 0cm 0cm}]{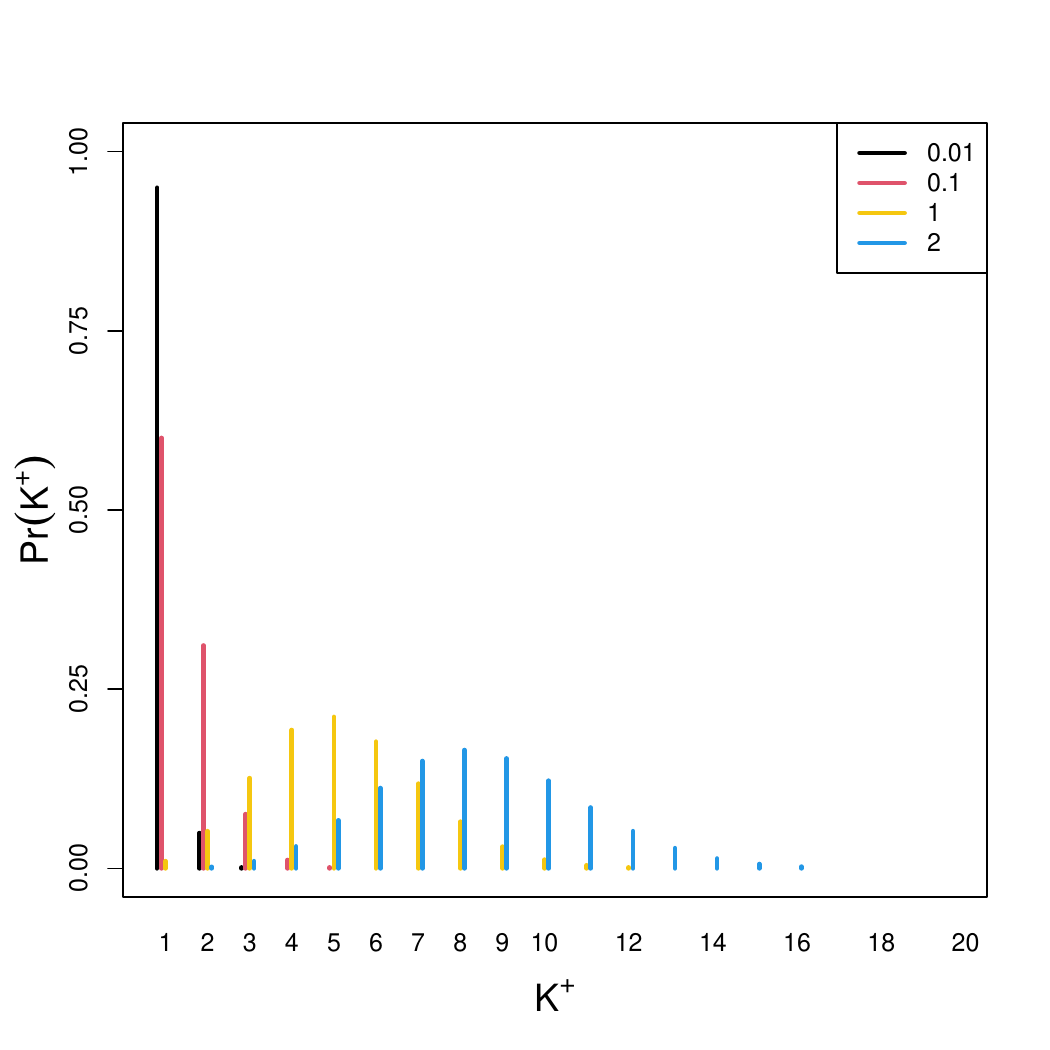}
\includegraphics[width=0.45\textwidth, trim={0 0cm 0cm 0cm}]{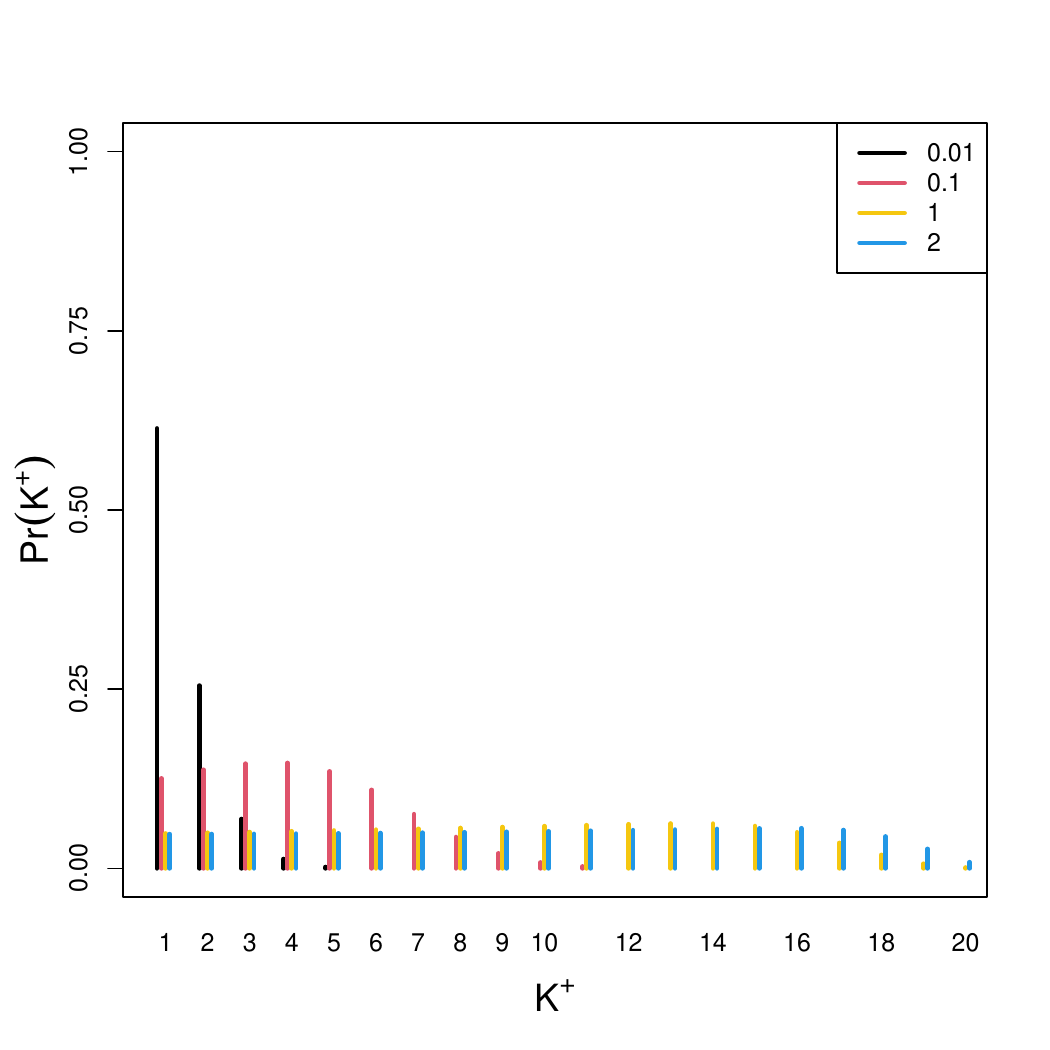}
\caption{Right panel: Induced prior distribution of $K^+$ for DPM ($K=\infty$) using Dirichlet$(\alpha)$ prior on the weights. Left panel: induced prior distribution of $K^+$ for MFMM using Uniform$(1,20)$ prior on $K$ and Dirichlet$(\alpha)$ prior on the weights. In both panels, different colors are associated to different values of the concentration parameter $\alpha \in \{0.01, 0.1, 1, 2\}$.
}
\label{fig:fipp.prior.kplus}
\end{figure}

The {\tt R} package \texttt{fipp} \citep{fipp} computes the implied prior on $K^+$ for three popular mixture models, namely a Dirichlet Process Model (DPM) and two versions of a MFMM that assume a symmetric Dirichlet prior on the weights with concentration parameter being fixed (static MFMM) or scaled by $K$ (dynamic MFMM). The implied prior on $K^+$ can be computed for any user-supplied prior on $K$ and the number of observations $n$. Figure \ref{fig:fipp.prior.kplus} shows the implied prior under the DPM and static MFMM for different values of the concentration Dirichlet parameter and $n=100$. By trial and error an expert user, having prior information on $K^+$, can tune the prior on $\bm{w}$ until the shape of the induced prior on $K^+$ resembles his/her prior belief on the number of occupied components. However, the user can only play a passive role in the sense that he/she cannot elicit the prior for $K^+$ in an intuitive way, for instance by eliciting it's mode or some other centrality parameter or by assigning a large probability mass to a specific range of values of $K^+$. Thus, although formally considering the induced prior on $K^+$ is certainly useful and improves the overall understanding of the FMM prior structure, the methods proposed don't permit users (particularly nonexperts) to  ``inform'' the FMM ($K^+$ in particular) in a straightforward way. Our contribution is to develop an approach that does this.  This is done by eliciting prior information through probabilistic statements associated with a user-supplied value of $K^+$.



Prior elicitation for $K^+$ can be challenging as it requires clearly defining what is meant by a  ``cluster'' and to clearly define the motivation behind using a FMM (\citealt{hennig:15}).  Once this has been established,  our approach of eliciting prior information follows the philosophy of the penalized complexity (PC) priors outlined in \cite{simpson_etal:2017}. In particular, we first specify a base or reference model and then the role of $\bm{w}$'s prior distribution is to ``shrink'' towards the reference model unless the data indicate otherwise. To do this we use an asymmetric Dirichlet distribution on the component weights. An appealing feature of this approach is that  scientific questions are able to guide reference model selection (e.g., a mixture with $K^+$ equal to a user-supplied value). 
Practitioners can then inform the FMM {\it a priori} by thinking directly about $K^+$ while tuning a parameter of the asymmetric Dirichlet, that controls {\it a priori} the FMM's sparseness (or lack thereof).

As with any Bayesian procedure, when the data  poorly inform a particular parameter, the prior can be highly influential on the resulting posterior distribution.  In this setting additional analysis must be executed to understand the exact impact the prior has on the posterior.  This is true for our prior construction for $K^+$.  However, our method is well suited to explore the prior's impact on the posterior $K^+$ because of the coherent way in which the prior is informed. As a result, it is very straightforward to carry out a sensitivity analysis and we provide one approach of doing this.

We finish the Introduction by briefly mentioning that the random probability measures that are commonly studied in the BNP literature (\citealt{MullerQuintana:2015, ghosal_van_der_vaart_2017}) set $K=\infty$ .  This essentially side-steps the need to formally consider $K$. That said, \cite{miller&harrison:2013} pointed out that estimating $K^+$ using a Dirichlet Process Model (DPM) can be problematic.   In fact, \cite{pmlr-v139-cai21a} find that estimating $K^+$ consistently depends on specifying components correctly (i.e., correctly defining the meaning of a cluster). As a result, \cite{lijoi_etal:2022}  studied in more depth the finite-dimensional Bayesian clustering from a normalized random measure with independent increments (\citealt{regazzini_etal:2022}) and \cite{ascolani_etal:2022} explored conditions necessary for a DPM to consistently estimate the $K^+$.  Recently, \cite{argiento:2022} and \cite{schnatter_etal:2021} made very interesting connections between BNP mixtures and FMMs while \cite{alamichel2023bayesian} studied the consistency in estimating $K^+$ (or lack there of) in a variety of mixture models.

The rest of the article is organized as follows.  In Section \ref{sec:background} we provide the necessary background for FMMs. Then in  Section \ref{sec:aFMM} we introduce the prior construction that permits informing FMMs and provide  some theoretical justifications.  Section \ref{sec:simulation} details a simulation study that compares our approach to a few other FMM procedures.  Section \ref{sec:applications} describes two applications. The first is the well known galaxy dataset and the second a biomechanics functional data example.  We end by providing some final comments in Section \ref{sec:discussion}.  All proofs and computational details are relegated to the online supplementary material along with additional details associated with the simulation study and applications detailed in Sections \ref{sec:simulation} and \ref{sec:applications}.

\section{Background on Bayesian Finite Mixture Models}\label{sec:background}

For computational purposes, the FMM in \eqref{fmm} is often re-expressed using latent component labels.  Doing so permits describing the model hierarchically.  To this end, let $z_1, \ldots, z_n$ denote $n$ component labels where $z_i = j$ implies that the $i$th unit belongs to the $j$th component.  Introducing component labels in the FMM and assuming each component density belongs to the same family permits expressing the FMM in \eqref{fmm} as
\begin{equation}
\begin{aligned} \label{eq:likelihood}
    y_i | z_i & \stackrel{ind}{\sim} f(\bm{\theta}_{z_i}), \ \mbox{ $i=1, \ldots, n$},  \\
    Pr(z_i = k | \bm{w}) & = w_k, \ \mbox{ $i=1, \ldots, n$}. 
\end{aligned}
\end{equation}
The Bayesian model is completed by assuming
\begin{align}
\bm{\theta}_k & \stackrel{iid}{\sim} \pi_{\theta}, \ \mbox{ $k=1, \ldots, K$},\label{eq:prior.theta} \\ 
    \bm{w} & \sim \mbox{Dirichlet}(\bm{\alpha}), \label{dirichlet_prior}
\end{align}
where $\pi_{\theta}$ denotes a prior distribution for $\bm{\theta}_k$ and $\mbox{Dirichlet}(\bm{\alpha})$ denotes a Dirichlet distribution with parameter $\bm{\alpha}$. As mentioned, it is possible to assign a prior distribution to $K$ also.  What we develop can be applied in that setting, but  we focus on the case when $K$ is fixed to a large value (\citealt{RousseauMengersen:2011}).

Even though $f(\cdot)$ and $\pi_{\theta}$ implicitly determine the type of clusters that are permitted in the FMM (e.g., spherical), their selection does not influence the implied prior on $K^+$.    Althernatively, the prior on $\bm{w}$ (and/or $\bm{\alpha}$) is directly connected to the implied prior on $K^+$.  However, both $f(\cdot)$ and $\pi_{\theta}$ and the prior on $\bm{w}$ impact the posterior distribution of $K^+$.  Thus any notion of posterior consistency associated with $K^+$ must necessarily consider both the type of clusters permitted based on $f(\cdot)$ and $\pi_{\theta}$ {\it and} the {\it a priori} number of clusters based on the prior for $\bm{w}$.  In this paper our focus is on informing the mixture through the implied prior on $K^+$ and hence focus on $\bm{w}$'s prior and assume that $f(\cdot)$ and $\pi_{\theta}$ are well specified.  

It is common to use a symmetric Dirichlet distribution as a prior for $\bm{w}$ so that $\bm{\alpha} = \alpha\bm{j}$ where $\bm{j}$ is a $K$-dimensional vector of 1s and $\alpha > 0$.   It is also quite common to fix $\alpha = 1/K$ since the resulting FMM would then approximate a Dirichlet process mixture (DPM) as $K \rightarrow \infty$ (\citealt{Ishwaran&James:2001}).   More recently, $\alpha$ has been assigned a prior.  In particular, \cite{walli:2016, schnatter:2019, greve:2020, schnatter_etal:2021} all assume $\alpha \sim Gamma(a, aK)$ so that $E(\alpha) = 1/K$.  The parameter $\alpha$ regulates the sparseness of the FMM in that as $\alpha \rightarrow 0$, $K^+$ decreases.

Although a symmetric Dirichlet prior for $\bm{w}$ is quite common, it is challenging to inform the induced prior on $K^+$ so that user specified values for $K^+$ are given large prior mass.   This results from the fact that the FMM is not explicitly parameterized in terms of $K^+$ and that the induced prior of $K^+$ is a function of $n$ and $K$ in addition to $\alpha$.  Next, we describe a prior construction that permits introducing expert opinion with regards to $K^+$ through an asymmetric Dirichlet prior on $\bm{w}$ which we will refer to as an asymmetric FMM (aFMM).  

\section{Asymmetric Dirichlet Finite Mixture Models} \label{sec:aFMM}
We desire to develop a method in which it is straightforward to guide the implied prior on $K^+$.  We do this by considering an asymmetric Dirichlet, defined below, as a prior for $\bm{w}$

\begin{definition} \label{def:asymDir}
The asymmetric Dirichlet, denoted by \mbox{Dirichlet}$(\bm{\alpha}_{1,2})$,  is a Dirichlet distribution with parameters $U, \alpha_1, \alpha_2$ such that $\bm{\alpha}_{1,2} =(\alpha_1\bm{j}_U, \alpha_2\bm{j}_{K-U})$ where $\bm{j}_U$ and $\bm{j}_{K-U}$ are $U$ and $K-U$ dimensional vectors filled with ones. 
\end{definition}

In Definition \ref{def:asymDir}, $U$ plays a crucial role as a user-supplied value on which the induce prior of $K^+$ is ``centered''.  An appealing property of our prior construction is that as $\alpha_1 \rightarrow \infty$ and $\alpha_2 \rightarrow 0$ prior mass concentrates on $U$ resulting in a mixture model with exactly $U$ occupied components.  We show this  in Proposition \ref{prop:prob_kp}, but first build some intuition why this is the case.  Note that under $\bm{w} \sim Dirichlet(\bm{\alpha}_{1,2})$
\begin{align}
  E(w_k)  & = \begin{cases*}
            \displaystyle\frac{\alpha_1}{\alpha_1U + \alpha_2(K-U)} &  $k=1,\ldots,U$ \\
            \displaystyle\frac{\alpha_2}{\alpha_1U + \alpha_2(K-U)} &  $k=U+1,\ldots,K$.
          \end{cases*} 
\end{align}
If $\alpha_1 \gg \alpha_2$ and $\alpha_2 \rightarrow 0$, then $E(w_k) \rightarrow 1/U$ for $k=1,\ldots,U$ and $E(w_k)  \rightarrow  0$, for $k=U+1, \ldots,K$.  Thus,  all prior mass is uniformly distributed over the first $U$ components, with no mass assigned to the remaining $K-U$ components. As a result, the implied prior on $K^+$ becomes a point mass at $U$ as $\alpha_1 \rightarrow \infty$ and $\alpha_2 \rightarrow 0$.  We show this more carefully in the following proposition the proof of which can be found in the supplementary material.  
\begin{proposition} \label{prop:prob_kp} Assume that $\bm{w} \sim Dirichlet(\bm{\alpha}_{1,2})$.  Then as $n \rightarrow \infty$ 
\begin{align}
\lim_{\alpha_1 \rightarrow \infty}\lim_{\alpha_2 \rightarrow 0} Pr(K^+ = U ~|~ K, n, \alpha_1, \alpha_2) = 1
\end{align}
\end{proposition}

\begin{figure}
\centering
\includegraphics[width=1\textwidth, trim={0 0cm 0cm 0cm}]{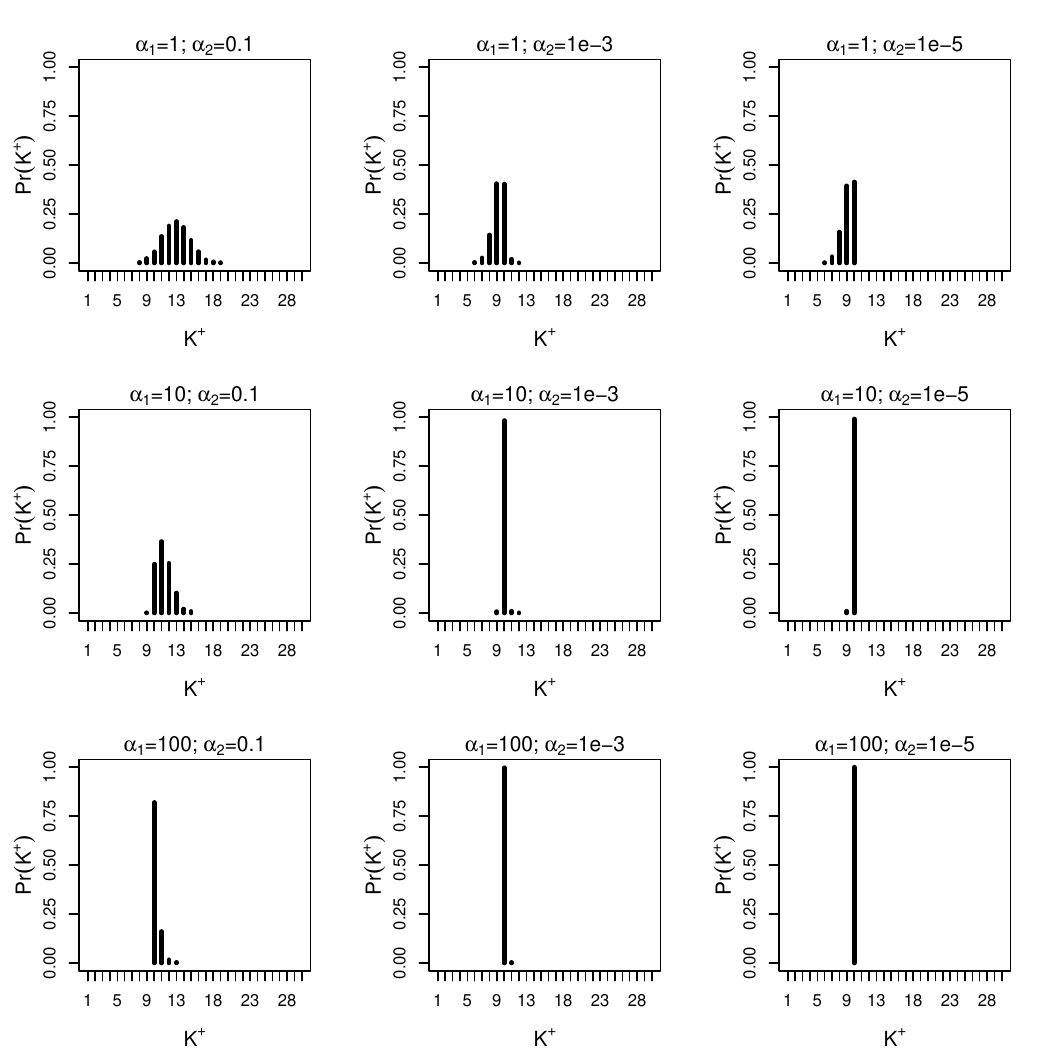}
\caption{Induced prior distribution of $K^+$ obtained via simulations, assuming the asymmetric Dirichlet prior with $\alpha_1=\{1,10,100\}$ and $ \alpha_2=\{0.1,0.001,0.00001\}$ and $U=10$, $n=100, K=30$. 
}
\label{fig:asymm.static}
\end{figure}

In order to visualize the implied prior of $K^+$ from the aFMM, we provide Figure \ref{fig:asymm.static} where different values for $\alpha_1$ and $\alpha_2$ are considered and $U=10$ which means the prior should be centered around $10$ non-empty components. The plots in Figure \ref{fig:asymm.static} are a graphical representation of the property of the aFMM expressed in Proposition \ref{prop:prob_kp}. Note that increasing $\alpha_1$ and decreasing $\alpha_2$ leads to an implied prior for $K^+$ highly concentrated on $U=10$; it is sufficient fixing $\alpha_1=U$ and $\alpha_2=0.001$ to get a spike on $K^+=10$ in this case. By moving $\alpha_1$ and $\alpha_2$ the probability mass can be moved to the left or right tails.  Doing so, the probability mass can be distributed either below or above $U$, in such a way that $U$ need not be connected to the center of the distribution. In particular, by decreasing $\alpha_1$ while keeping $\alpha_2$ small (e.g. smaller than $0.001$), we get more probability in the left tail hence  $U=10$ can be intended as a \emph{soft} upper bound. Analogously, by increasing $\alpha_2$ while $\alpha_1$ being large, we obtain a prior with heavy right tail hence $U=10$ can be intended as a \emph{soft} lower bound.  

The aFMM is constructed in such a way that the implications of the theory in \cite{RousseauMengersen:2011} hold.  We state this carefully in the following remark.   
\begin{remark}\label{remark:mengerson&R}
    The  aFMM as described in \eqref{eq:likelihood} - \eqref{eq:prior.theta} with $\bm{w} \sim \mbox{Dirichlet}(\bm{\alpha_{1,2}})$ satisfies the assumptions in  \cite{RousseauMengersen:2011} so that if $\min(\alpha_1, \alpha_2) < d/2$ the asymmptotic vanishing weights property holds. 
\end{remark}
\noindent The proof of Remark \ref{remark:mengerson&R} follows from the arguments laid out in \cite{alamichel2023bayesian}.  Note that Remark \ref{remark:mengerson&R} holds only if $\alpha_1$ and $\alpha_2$ are considered fixed quantities. 

\subsection{Prior Distributions for $\alpha_1$ and $\alpha_2$}

There are a number of ways to treat ($\alpha_1$, $\alpha_2$).  The list includes A) Fix $\alpha_1$ and $\alpha_2$ to prespecified values, which corresponds to an asymmetric case of the static model described in \cite{schnatter_etal:2021}; B) assume that $\alpha_1$ is unknown with an assigned prior distribution and fix $\alpha_2$ to a small value; C) Fix $\alpha_1$ to a prespecified value and assume $\alpha_2$ is unknown with an assigned prior distribution; and D) assume both $\alpha_1$ and $\alpha_2$ are unknown and assign both a prior distribution.   Each possibility results in a specific balance between computation cost and model flexibility.  In what follows, we focus primarily on the case that $\alpha_1$ is assigned a prior and $\alpha_2$ is fixed at some prespecified value.  This permits us to treat $U$ as an soft upper bound for $K^+$ with $\alpha_2$ controlling the rigidity of the upper bound in the sense that as $\alpha_2 \rightarrow 0$, then $U$ becomes a hard upper bound.  There are a number of prior distributions that might be considered for $\alpha_1$.  For reasons we provide shortly, we focus primarily on a prior that has connections to PC priors (\citealt{simpson_etal:2017}).

\subsection{Penalized Complexity Motivated Prior}



Our prior construction is guided by a key idea on which PC priors are based.  Mainly, the prior is treated as a mechanism that regulates the behaviour of the model with respect to a parsimonious version of it called the {\it base model}. A PC prior guarantees that the base model is favored unless data support an alternative model. Typically the alternative model is assumed to be more flexible (or complex) than the base one, or an over-parameterized version of the base one. The PC prior is formally defined as an exponential distribution on a measurement scale quantifying the increased complexity of the alternative (i.e. flexible) model with respect to the base one, where {\it complexity} is measured by the Kullback–Leibler divergence (KLD,  \cite{kld-1951}). A brief review of the principles and the practical steps underlying the construction of PC priors, as originally proposed in \cite{simpson_etal:2017}, is in supplementary material.

Our aim is to construct a prior for the asymmetric Dirichlet parameters $(\alpha_1,\alpha_2)$ such that the induced prior on $K^+$ guarantees that a mixture model with a user-defined number of non-empty components, say $U \in[1,K]$, is favoured unless data support an alternative FMM. Thus, the implied prior for $K^+$ is used as a mechanism to regulate the behaviour of the FMM with respect to a {\it base finite mixture model} (base FMM). In general, the base FMM is a FMM with  $K^+=U$, for some $U$ selected by users according to the goal of the analysis and/or their prior knowledge about $K^+$. 
A base FMM favouring $K^+=U$ can be obtained by treating the asymmetric Dirichlet from Definition \ref{def:asymDir}, $\mbox{Dirichlet}(\bm{\alpha}_{01,02})$, as a prior on $\bm{w}$, with parameters $\alpha_{01}=\infty$, $\alpha_{02}=0$, and $U$. Note, we will use $\bm{\alpha}_{01,02}$ to refer to the Dirichlet parameters under the base model. Because the asymmetric Dirichlet is not defined for $\alpha_{1} = \infty$ and $\alpha_{2} = 0$, as a {\it practical} base FMM we will use a ``large'' value for $\alpha_{01}$ and a ``small'' value for $\alpha_{02}$. Numerical experiments lead us to set $\alpha_{01}=U$ and $\alpha_{02}=10^{-5}$. Thus, $\mbox{Dirichlet}(\bm{\alpha}_{01,02})$, with $\alpha_{01}=U, \alpha_{02}=10^{-5}$ and for a specific $U$ is our  ``practical base model'' in general situations where we want a mixture model favouring $U$ clusters (this choice worked well for a variety of values for $n$ and $K$).



Constructing the PC prior requires quantifying how much a FMM  with parameters $(\alpha_1,\alpha_2)$ {\it deviates} from the particular base FMM. Let $\bm{g} \sim \mbox{Dirichlet}(\bm{\alpha}_{1,2} )$ be the asymmetric Dirichlet under the base FMM with parameters $\alpha_1=\alpha_{01}$, $\alpha_2=\alpha_{02}$ and $U$, while $\bm{p} \sim \mbox{Dirichlet}(\bm{\alpha}_{1,2})$ be the asymmetric Dirichlet under the alternative FMM with parameters $\alpha_1>0$, $\alpha_2>0$ and $U$. (Note that $\bm g$ and $\bm p$ have  different values of the parameters $\alpha_1$ and $\alpha_2$ but the same $U$). The {\it deviation} from the alternative FMM to the base FMM is measured using the KLD between $\bm{p}$ and $\bm{g}$ 
\begin{align}
    KLD(\bm{p} || \bm{g}) & = \log\Gamma(\alpha_1 U + \alpha_2(K-U)) - \log\Gamma(\alpha_{01}U + \alpha_{02}(K-U)) - \nonumber \\ 
    & ~~~ (U\log\Gamma(\alpha_1) + (K-U)\log\Gamma(\alpha_2)) + U\log\Gamma(\alpha_{01}) + (K-U)\log\Gamma(\alpha_{02}) +   \nonumber\\
    & ~~~  U(\alpha_1-\alpha_{01})[\psi(\alpha_1) - \psi(\alpha_1 U+ \alpha_2(K-U))] +  \nonumber\\ 
    & ~~~ (K-U)(\alpha_2-\alpha_{02})[\psi(\alpha_2) - \psi(\alpha_1 U + \alpha_2(K-U))],
\label{eq:kld}
\end{align}
where $\Gamma$ and $\psi$ are, respectively, the gamma and  digamma functions.

The function in Eq. (\ref{eq:kld}) depends on the asymmetric Dirichlet parameters $(\alpha_{01},\alpha_{02})$ under the practical base model, the asymmetric Dirichlet parameters $(\alpha_1,\alpha_2)$ under the alternative model, and the user-supplied value for $U$. Function (\ref{eq:kld}) 
represents a suitable scale to measure deviations from the base FMM. 


For ease of interpretation (\ref{eq:kld}) is transformed to a unidirectional distance measure $d(\alpha_1, \alpha_2)=  d(p || g)=\sqrt{2\text{KLD}(\bm{p} || \bm{g})}$. (Note our notation for $d$ focuses on a two-dimensional function of the Dirichlet parameters $(\alpha_1,\alpha_2)$ because these are the parameters we need to assign a prior to, but $d$ also depends on the user-supplied $U$ and the choice of the practical base model $(\alpha_{01},\alpha_{02})$). When $d=0$, the FMM corresponds to the base FMM, i.e., a FMM favouring $K^+=U$ non-empty components. As $d$ increases the FMM is allowed to deviate from the base FMM, with deviations occurring either as a FMM favouring $K^+ < U$ (i.e. sparser mixture) or $K^+ > U$ (i.e. less sparse mixture).

\subsubsection{PC prior for $\alpha_1$ conditional on $\alpha_2$ being fixed}

Following \cite{simpson_etal:2017} we consider an exponential distribution on $d(\alpha_1, \alpha_2)$, with rate $\lambda>0$, 
so that the mode is always at the base model $d=0$, or $K^+=U$, and the penalization rate is constant. However, in our case the distance $d(\alpha_1,\alpha_2)$ is a surface that varies over $\alpha_1$ and $\alpha_2$ and potentially one may consider two parameters $\lambda_1$ and $\lambda_2$ to penalize deviations along $\alpha_1, \alpha_2$ at different rates. 

\begin{figure}
\centering
\includegraphics[width=0.48\textwidth, trim={0 0cm 0cm 0cm}]{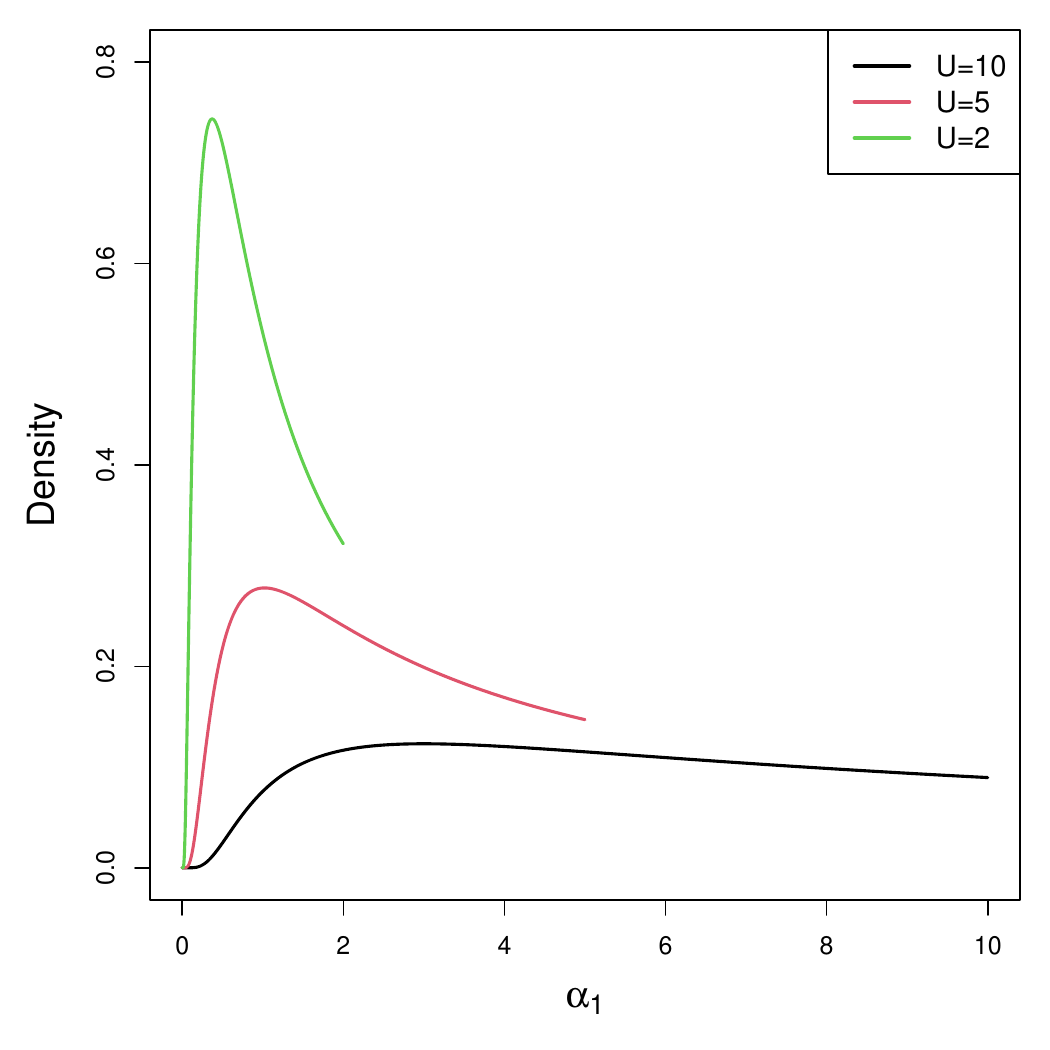}
\includegraphics[width=0.48\textwidth, trim={0 0cm 0cm 0cm}]{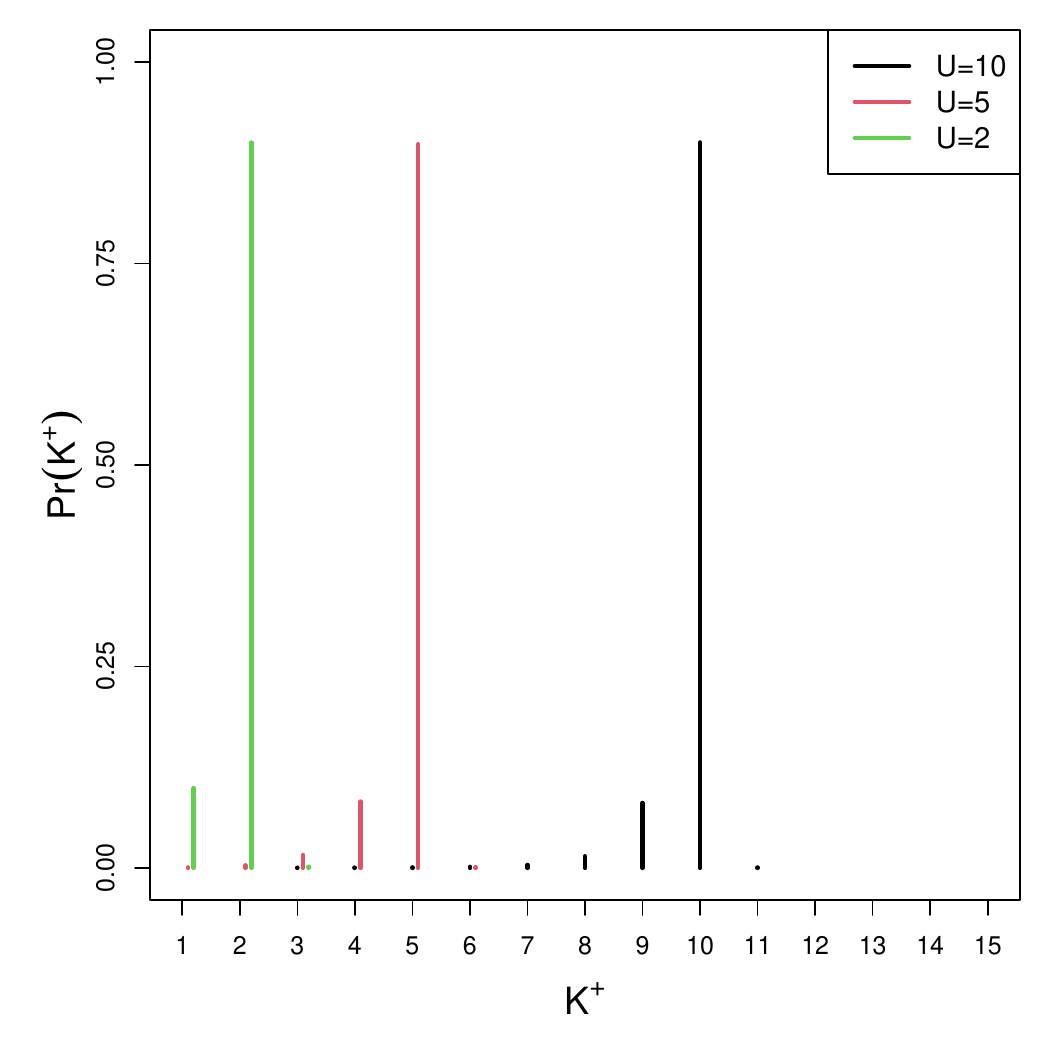}
\caption{PC prior on $\alpha_1 \in (0,U)$ given $\alpha_2=1e-5$ (left) and the implied prior on $K^+$ (right), for three different choices of $U$ and tail probability $tp=0.1$. 
}
\label{fig:example_pc}
\end{figure}

\begin{figure}
\centering
\includegraphics[width=0.48\textwidth, trim={0 0cm 0cm 0cm}]{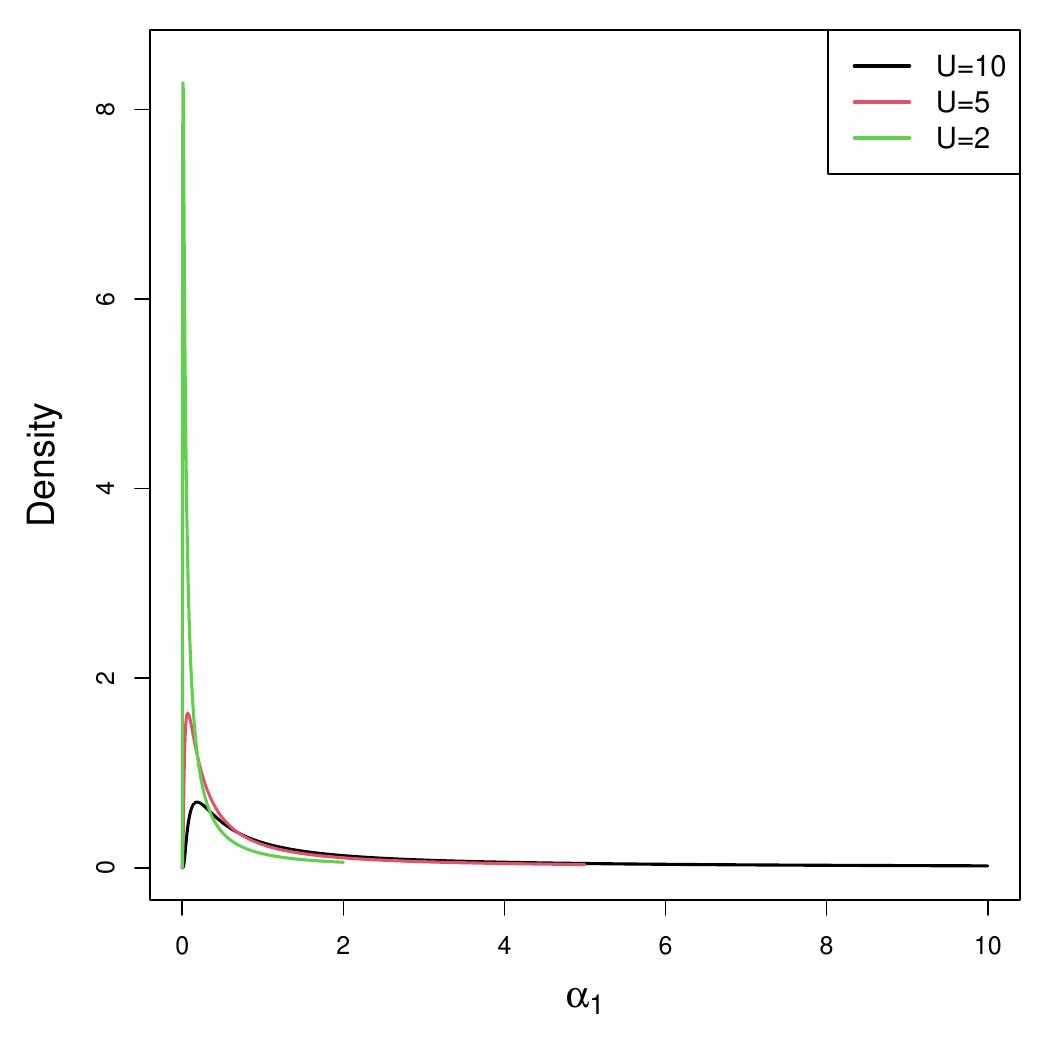}
\includegraphics[width=0.48\textwidth, trim={0 0cm 0cm 0cm}]{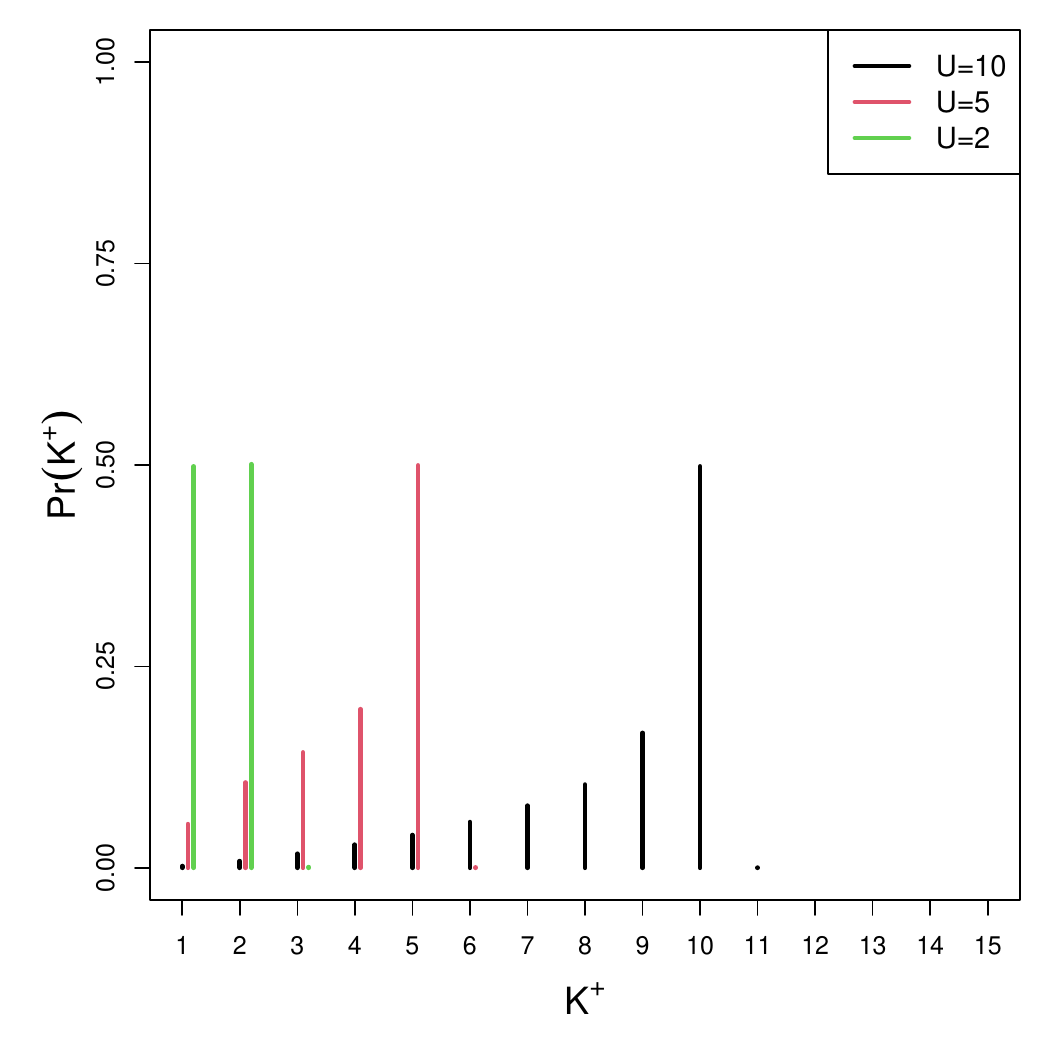}
\caption{PC prior on $\alpha_1 \in (0,U)$ given $\alpha_2=1e-5$ (left) and the implied prior on $K^+$ (right), for three different choices of $U$ and tail probability $tp=0.5$. 
}
\label{fig:example_pc_05}
\end{figure}

From Figure S1 of the supplementary material we can visually inspect the KLD in (\ref{eq:kld}) and we see that it varies more sharply along $\alpha_1$ than  $\alpha_2$. An exponential prior on $d(\alpha_1,\alpha_2)$ with distinct decay rates $\lambda_1$ and $\lambda_2$ will permit the user to get an induced prior on $K^+$ with mode at $U$ and, at the same time, the ability to tune the probability mass assigned to $K^+<U$ and $K^+>U$ independently, by careful selection of  $\lambda_1$ and $\lambda_2$. This strategy is useful when users have precise information about $K^+$ and wish to center the prior for $K^+$ on $U$. From a computational point of view this strategy is quite expensive as numerically deriving the prior implies optimizing over two decay rate parameters which will slow the MCMC considerably. 

We seek a simpler solution here in the form of a conditional PC prior on $\alpha_1 \in (0,U]$, given $\alpha_2$ set to a small value. Doing so, the user-supplied $U$ can be intended to be an upper bound for $K^+$ which we believe works well in many applications. 
From our experiments $\alpha_2=10^{-5}$ is small enough to have $Pr(K^+ > U)$ approximately zero, with $U$ playing the role of a {\it lenient} upper bound; 
however, by increasing $\alpha_2$  the right tail probability will increase too, making $U$ a softer upper bound. 

The PC prior on $\alpha_1$ conditional on $\alpha_2=10^{-5}$ is the (truncated) exponential prior on the distance $d(\alpha_1,\alpha_2=10^{-5})$, 
and follows by a change of variable transformation:
\begin{equation}
\pi(\alpha_1) = \frac{\lambda \exp\left(-\lambda d(\alpha_1,\alpha_2=10^{-5})\right)|d'(\alpha_1,\alpha_2=10^{-5})|}{1-\exp(\lambda d(\alpha_1=U,\alpha_2=10^{-5}))}, \quad \lambda>0, \quad 0<\alpha_1 \leq U
\label{eq:pc.alpha1}
\end{equation}
Details on the numerical derivation of (\ref{eq:pc.alpha1}) can be found in supplementary material S3. One appealing feature of (\ref{eq:pc.alpha1}) is that the user is only required to handle a single decay rate parameter $\lambda$, hence the scaling of the PC prior according to the user prior guess about $K^+$ greatly simplifies. To ``scale the PC prior'' for us means to choose $\lambda$ in Eq. (\ref{eq:pc.alpha1}).  

Scaling the PC prior can be approached from the following situations: either the user might have information on the maximum number of clusters possibly present in the data at hand, or on the number of clusters that he/she is able to interpret.  We propose computing $\lambda$ based on a user-defined probabilistic statement like 
\begin{equation}
    Pr(K^+ < U) = tp
\end{equation}
In other words, our aim is to help the user select the $\lambda$ that corresponds to assigning a certain probability, denoted as  $tp$, to the left-tail $(1,U-1)$. We use simulations to find the optimal $\lambda$ that realizes a left-tail probability equal to the user-defined $tp$; the procedure is described in supplementary material. Figures \ref{fig:example_pc} and \ref{fig:example_pc_05} display the implied prior on $K^+$ obtained by setting the left-tail probability equal to $0.1$ and $0.5$, respectively, and different values of the lenient upper bound $U$.

A general appealing property of the PC prior is that it accommodates the user-selected value of $K$ and the number of observations $n$ in the case study at hand in a natural way. There are two reason why this is the case. The prior in Eq. (\ref{eq:pc.alpha1}) derives from assuming an exponential prior on the KLD (which depends on $K$), hence it ``adapts'' automatically to any value of $K$ the user may choose. In addition, the simulation-based algorithm to numerically derive the prior for $K^+$ requires $n$ as an input, other than $K$, thus the number of observations in the application at hand would be automatically taken into account in the (induced) prior for $K^+$. 



\subsection{Special Cases}

The aFMM has as special cases other commonly used over-parameterized FMMs.  For example, if we set $U = 1$,  then the asymmetric Dirichlet prior can induce sparsity in the sense that $K^+$ is much smaller than $K$. This aFMM would have shrinkage properties similar to that of  sFMM as described in the following remark
\begin{remark}\label{remark:sparseFMM}
    If $U=1$ and $\alpha_2$ is fixed at a small value then  as $\alpha_1 \rightarrow 0$ the $Pr(K^+ = 1 ~|~ K, n, \alpha_1, \alpha_2) = 1$ resulting in a sFMM. 
\end{remark}
\noindent The proof of remark \ref{remark:sparseFMM} follows from arguments similar to those found in the proof of Proposition \ref{prop:prob_kp}.

Shrinkage properties similar to sFMM can also be achieved through $tp$.  When $tp$ is set to a large value (i.e., close to one) then the induced prior on $K^+$ will be such that the majority of prior mass is concentrated on values (much) smaller than $U$ (when $\alpha_2$ is small).  
Finally, setting $U=0$ recovers the symmetric Dirichlet prior for $\bm{w}$ with $\alpha_2$ acting as the lone concentration parameter. As a result, all FMM methods that have been developed using a symmetric Dirichlet prior can be employed. 

\section{Simulation Study}\label{sec:simulation}
\begin{figure}
  \centering
  \includegraphics[page=4, scale=0.8]{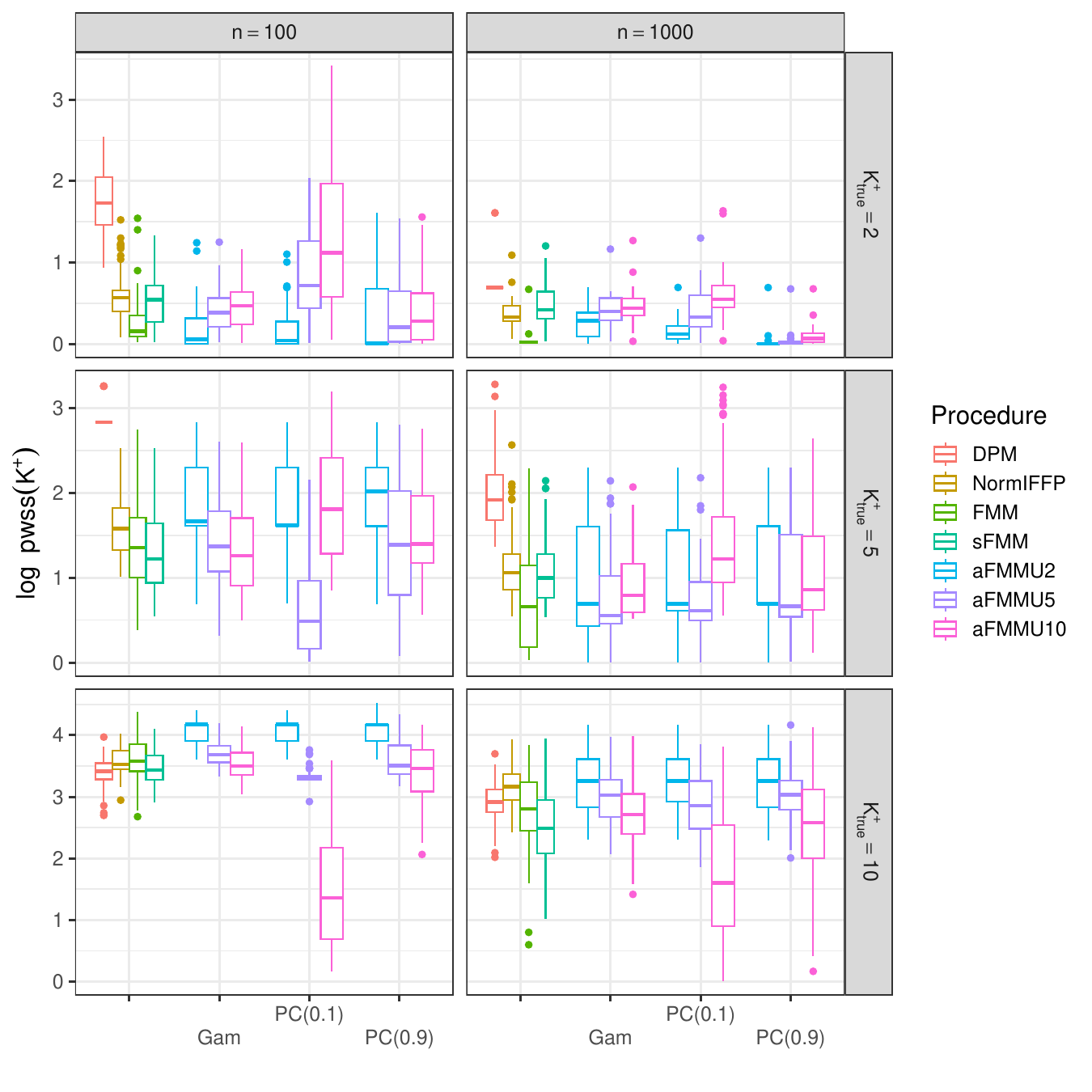}
\caption{$\log(\mbox{pwss}(K^+) + 1)$ for {\it Data Type 2}. Each row corresponds to results associated with the value of $K^+$ used to generate data.}  \label{fig:asymSimStudy.var.data2}
\end{figure}

\begin{figure}
  \centering
  \includegraphics[page=10, scale=0.75]{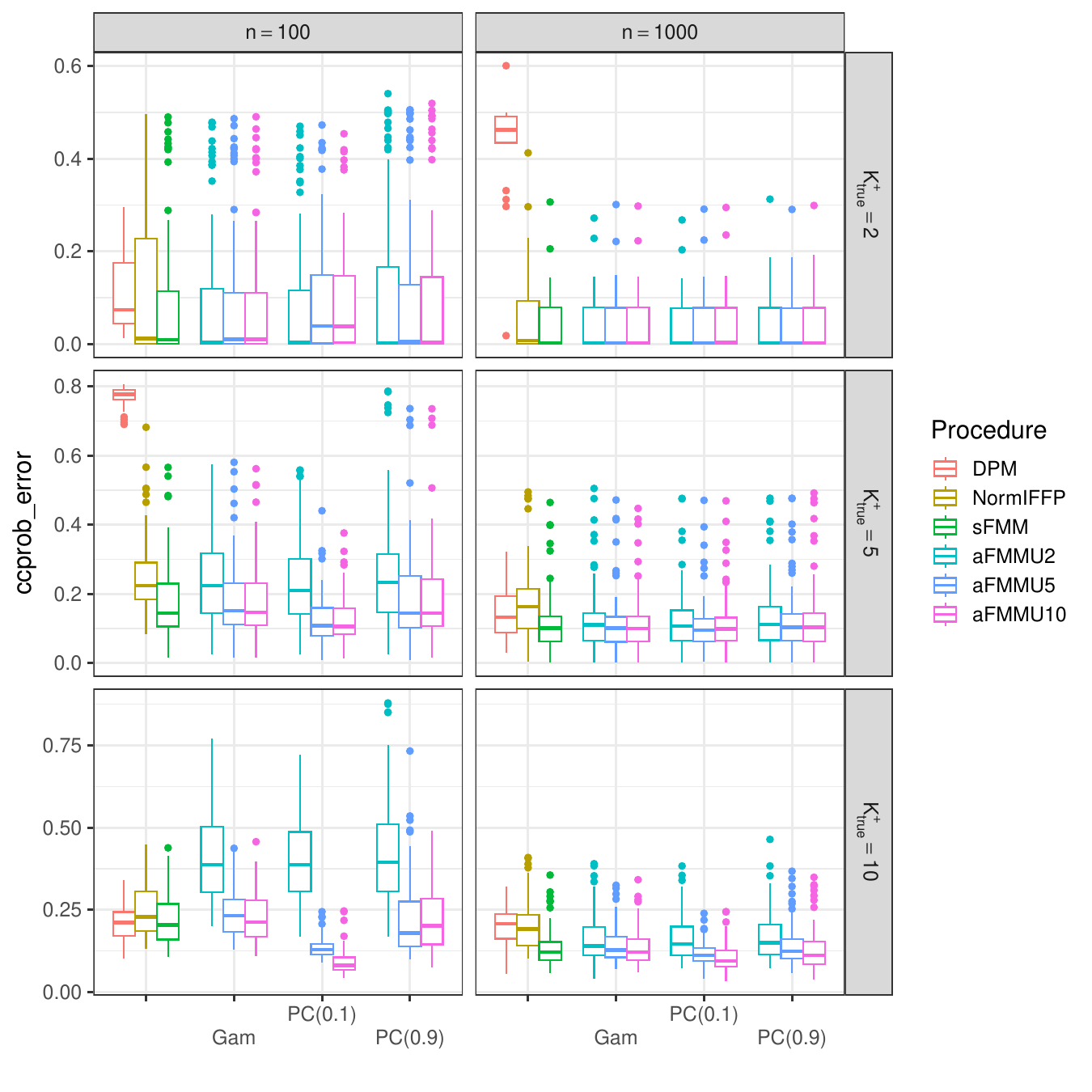}
    \caption{Error of co-clustering probabilities for {\it Data Type 2}. Each row corresponds to results associated with the value of $K^+$ used to generate data.}
   \label{fig:asymSimStudy.ccprob.data2}
\end{figure}

In order to illustrate the aFMM's performance in estimating $K^+$, we conduct a numerical experiment.  Even though an asymmetric prior distribution on $\bm{w}$ (and as a result an informed prior for $K^+$) can be employed for any FMM, in the simulation and application that follow we focus on the case that $f_k(\cdot)$ is a Gaussian.  As a result, \eqref{fmm} becomes
\begin{align} \label{eq:gaussian.mixture}
y_i \sim \sum_{k=1}^K w_k \mbox{N}(\mu_k, \sigma^2_k)
\end{align}
so that $\bm{\theta}_k = (\mu_k, \sigma^2_k)$.   After introducing the component labels, the augmented data model becomes
\begin{equation}
\begin{aligned} \label{eq:gaussian.likelihood}
    y_i ~|~ z_i & \sim \mbox{N}(\mu_{z_i}, \sigma^2_{z_i})  \\
    Pr(z_i = k ~|~ \bm{w}) & = w_k,
\end{aligned}
\end{equation}
and we use the following prior distributions
\begin{equation}
\begin{aligned} \label{eq:prior}
    \mu_k & \sim \mbox{N}(\mu_0, \sigma^2_0) \\
    \sigma^2_k & \sim \mbox{Inverse-Gamma}(a_0,b_0) \\
    \bm{w} & \sim \mbox{Dirichlet}(\bm{\alpha}_{1,2})\\
    \alpha_1 ~|~ \alpha_2, U & \sim PC(U, tp).  
\end{aligned}
\end{equation}
Here ``Inverse-Gamma'' denotes an inverse Gamma distribution parameterized so that the prior mean of $\sigma^2_k$ is $b_0/(a_0-1)$.  For hyper-prior values we set $\mu_0 = \mbox{mean}(\bm{y})$, $\sigma^2_0 = 10^2$ which correspond to one of the prior specifications that was employed in \cite{Bettina:2021}.  We also set $a_0 = 3$ and $b_0= 2$ which is also very similar to one of the prior specifications in \cite{Bettina:2021}. We set $K=25$ in all our implementations of the aFMM.  All computation is carried out using the {\tt informed\_mixture} function that can be found in the {\tt miscPack} {\tt R}-package that is available at {\tt https://github.com/gpage2990}.   Data sets are generate using \eqref{eq:gaussian.mixture} as a data generating mechanism in the following two ways: 
\begin{enumerate}
    \item[] {\bf Data Type 1}: Set $K=K^+$ for  $K^+ \in \{2, 5, 10\}$ and then generate $n \in \{100, 1000\}$ observations by setting $\bm{w} = 1/K\bm{j}$, $\sigma_k = 0.5$, and $\mu_k = 3(k-1)$.  In this scenario there are always exactly $K^+ \in \{2, 5, 10\}$ clusters with centers displaying little overlap.  As $n$ increases from 100 to 1000 the number of observations in each component increases but is still quite uniform across the $K^+$ clusters.  
    \item[] {\bf Data Type 2}: Use \eqref{eq:gaussian.likelihood} - \eqref{eq:prior} to generate $n \in \{100, 1000\}$ observations by setting $K=25$, $\alpha_1 = U$, $\alpha_2=10^{-3}$, $A=1$, $\mu_0 = 0$, $\sigma^2_0 = 3$, and $U \in \{2, 5, 10\}$.  In this scenario  clusters may not be well separated and the number of observations in each of the $K^+$ clusters can vary greatly.  As a result, this data generating scenario can be much more challenging than the first in estimating $K^+$.      
\end{enumerate}
Examples data sets created using the procedure just described for {\it Data Type 1} and {\it Data Type 2} are provided in Figures S2 and S3 of the supplementary material.  For each data type 100 datasets are generated and to each we fit an aFMM for $U \in \{2, 5, 10\}$ under the following prior specifications
\begin{enumerate}[align=left]
\item[{\bf Gam}:] Fix $\alpha_2 = 10^{-5}$ and use $\alpha_1 \sim Gamma(a,b)$ where $a=10$ and $b=(10U)^{-1}$,
\item[{\bf PC(0.1)}:] Fix $\alpha_2 = 10^{-5}$ and use $\alpha_1 ~|~ \alpha_2 \sim PC(U, tp=0.1)$, which means that the user prior statement is $Pr(K^+ < U)=0.1$. 
\item[{\bf PC(0.9)}:] Fix $\alpha_2 = 10^{-5}$ and use $\alpha_1 ~|~ \alpha_2 \sim PC(U, tp=0.9)$, which means that the user prior statement is $Pr(K^+ < U)=0.9$. 
\end{enumerate}
\noindent Each of these prior specifications are found on the $x$-axis in Figures \ref{fig:asymSimStudy.var.data2} and \ref{fig:asymSimStudy.ccprob.data2}.  In addition to fitting the aforementioned aFMM models to each dataset, for context,  we also fit the following methods:
\begin{enumerate}[align=left]
\item[{\bf sFMM}:] sparse FMM as described in \cite{walli:2016} such that $\alpha \sim Gamma(10,10K)$,
\item[{\bf FMM}:] A FFM fit using RJMCMC as employed in the {\tt mixAK} {\tt R}-package (\citealt{mixAK-Rpackage}),
\item[{\bf DPM}:] A Dirichlet Process Mixture model with dispersion parameter set at 1,
\item[{\bf NormIFPP}:] The normalized independent finite point process FMM described in \cite{argiento:2022} and fit using the {\tt AntMAN}  {\tt R}-package (\citealt{AntMAN-Rpackage}).  
\end{enumerate}

\noindent Hyper-prior values for all methods listed were selected to match as much as possible those used in the aFMM procedures.  For the NormIFPP method we employed values that were suggested in \cite{argiento:2022}.  The simulation was executed using GNU parallel (\citealt{tange_2022_6891516}).

To compare each methods ability to estimate $K^+$ we recorded the bias associated with the posterior mode of $K^+$ and two other metrics that evaluate the accuracy of the entire posterior distribution of $K^+$.  The first is a posterior probability weighted sum of squares associated with $K^+$ as defined below
\begin{align} \label{eq:varKp}
    \mbox{pwss}(K^+) \stackrel{\mbox{def.}}{=} \sum_{k=1}^{K} (k - K^+_{\mbox{true}})^2Pr(K^+ = k ~|~ \bm{y}),
\end{align}
where $K^+_{\mbox{true}}$ denotes the value of $K^+$ used to generate the data. This metric takes into account both the spread and location of the posterior distribution of $K^+$ relative to $K^+_{\mbox{true}}$ with smaller values indicating a more precise estimate of $K^+$.  The second metric evaluates the accuracy of the co-clustering probabilities for each observation and is defined as follows
\begin{align*}
\mbox{ccprob\_error} \stackrel{\mbox{def.}}{=} \sum_{j=1}^n \sum_{\ell < j}(I[j\sim\ell] - Pr(z_j = z_{\ell} ~|~ \bm{y} ))^2, 
\end{align*}
where $I[j\sim\ell]=1$ if unit $j$ and $\ell$ belong to the same cluster and zero otherwise.  Small values of {\tt ccprob\_error} indicate more accurate estimation of the underlying partition and as a result, the number of clusters. In Figures \ref{fig:asymSimStudy.var.data2} - \ref{fig:asymSimStudy.ccprob.data2} we report results for {\tt ccprob\_error} and $\mbox{pwss}(K^+)$ under {\it Data Type 2}.  Results associated with with bias and from {\it Data Type 1} are provided in Figures S4-S7 of the supplementary material.

From Figure \ref{fig:asymSimStudy.var.data2} note that when $U = K^+_{\mbox{true}}$ the aFMM performs the best regardless of prior on $\alpha_1$ except for FMM that in some scenarios (i.e., for some choice of $tp$) performs better than aFMM even when $U = K^+_{\mbox{true}}$.  This is unsurprising. When $U \ne K^+_{\mbox{true}}$,  aFMM continues to perform competitively (at least one aFMM procedure is the best or second best) in all scenarios while the competing methods do well in some scenarios but poorly in others.   Note further how the ``sparse'' aFMM ($PC(0.9)$) tends to perform well when $K^+_{\mbox{true}} < U$.  As expected, setting $U$ to a value that is far from $K^{+}_{\mbox{true}}$ tends to result in poor performance for the aFMM.   Trends for {\it Data Type 1} are similar (see Figure  S5 of the supplementary material).  

Regarding Figure \ref{fig:asymSimStudy.ccprob.data2}, first note that the {\tt FMM} procedure is not included.  This is due to the fact that it is not possible to compute the co-clustering probabilities based on output provided by the {\tt mixAK} package.  Now, it appears that all methods save the DPM perform similarly when $K^{+}_{\mbox{true}} = 2$ regardless of sample size. For $K^{+}_{\mbox{true}} = 5$ it appears that the aFMM performs best when $n=100$ so long as $U>2$ and for $n=1000$ the aFMM performs similarly to sFMM while outperforming DPM and NormalIFPP regardless of the prior on $\alpha_1$. However, when $K^{+}_{\mbox{true}}=10$ it appears that aFMM performs best when $U > 2$ and one of the two PC priors are employed.   The upshot of the simulation study is that the estimation $K^+$ under the aFMM  performs well if $U$ is not far from the truth for small $n$ and performs very competitively when $n$ is large regardless of $U$. 


\section{Applications}\label{sec:applications}
In this section we further illustrate the utility of the aFMM in two applications.  The first is the well known galaxy dataset while the second is a biomechanic application where prior information is elicited from exercise scientists.  We consider the galaxy data to illustrate the use of our prior construction as a principled tool to evaluate the sensitivity of the clustering configuration with regards to the induced prior on $K^+$ . The biomechanic data permits illustrating how our prior construction can be intuitively employed to accommodate prior beliefs elicited from experts that approach an analysis from different perspectives. The biomechanic  data will be modelled from a functional data perspective.  This  requires a model that is more complex than that described in \eqref{eq:gaussian.mixture} - \eqref{eq:prior} which we detail in \ref{sec:fdamodel}. 

\subsection{Galaxy Data} \label{sec:galaxy} 
The well known galaxy dataset (available in the \texttt{MASS} library) contains the velocities (km/sec) of 82 galaxies. 
This dataset has been widely used to illustrate methods in the clustering literature (\citealt{Bettina:2021}). \cite{Aitkin:2001} argues that there are 3 clusters if equal variance components are assumed and 4 if variances are allowed to be unequal. Others claim that there are more than 4 clusters (ranging between 6 and 9 (\citealt{Bettina:2021})).  
Due to the uncertainty associated with $K^+$ in the galaxy data, they are well suited to illustrate how our prior construction can be used to carry out a principled sensitivity analysis for $K^+$. This is done by fitting an aFMM for a sequence of $U$ values and then exploring the prior's impact on properties of the mixture model like model-fit and co-clustering probabilities. 
With this in mind, we fit the aFMM to the galaxy data for $U \in \{2, \ldots, 10\}$, $tp \in \{0.1, 0.5\}$ and $\alpha_2 = 10^{-5}$. We employ the same prior distribution specification as in Section \ref{sec:simulation}.    The aFMM is fit by collecting 1000 MCMC samples after discarding the first 10,000 as burn-in and thinning by 100 (i.e., 110,000 total MCMC samples). 

\begin{figure}
  \centering
  \includegraphics[scale=0.7]{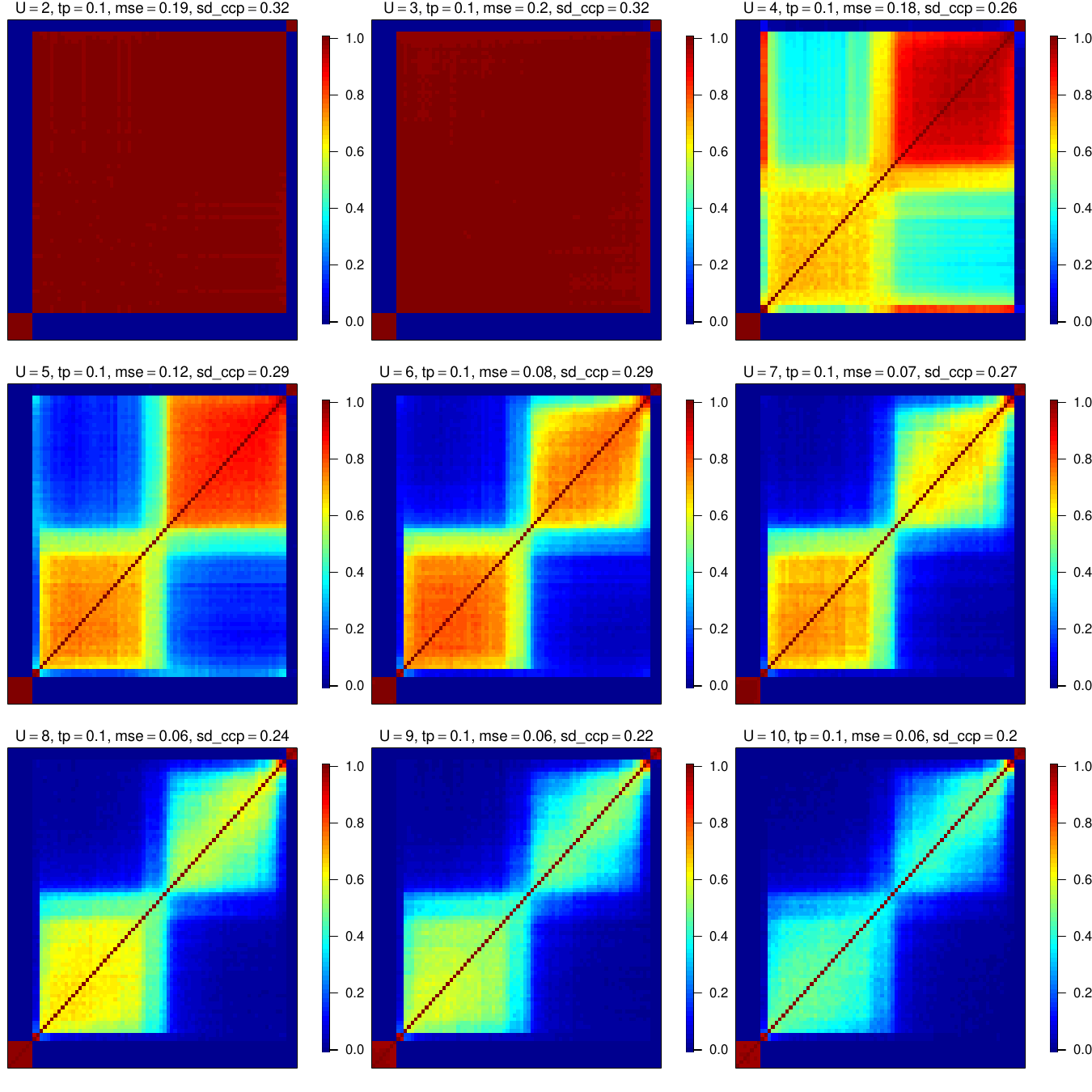}
    \caption{Posterior co-clustering probabilities for the Galaxy data from an aFMM fit using $U \in \{2, \ldots, 10\}$ and $tp = 0.1$. In addition,  sd\_ccp as defined in \eqref{eq:sdccp} and mse as defined in \eqref{eq:mse} are provided. In middle row panels, choice of $U \in \{5,6,7\}$ balances best the quality of cluster configuration and model fit (high purity and moderate mse).}\label{fig:galaxy_ccp}
\end{figure}

The posterior distributions of $K^+$ for $U \in \{3, 5, 7, 10\}$ and the induced priors on $K^+$ are provided in Figure S8.  Notice that the posterior distribution of $K^+$ is influenced quite heavily by $U$ for $tp = 0.1$ and also for $tp = 0.5$, but less so.  In both cases $mode(K^+ ~|~ \bm{y}) = U$ for each value of $U$.  At first glance this may seem problematic, but $U$'s impact on the $mode(K^+ ~|~ \bm{y})$ is not seen in the clustering configuration for $U > 4$.  To see this, we provide the co-clustering probability matrices for $tp = 0.1$ in Figure \ref{fig:galaxy_ccp}.  The rows and columns of the co-clustering matrices are ordered by velocity.  Notice that for $U \le3$ there are three clear clusters with little movement between them.  This is expected as in the galaxy data there are three groups of velocities that are well separated (see Figure S9 for density estimates).  For  $U \ge 6$ there appear to be six clusters, but the co-clustering probabilities among units that belong to the two big clusters decrease as $U$ increases.  Thus, even though  $mode(K^+ ~|~ \bm{y})$ based on the aFMM follows $U$ for these data, it does so not by forming clusters that don't exist but by grouping units within the two big clusters in a fairly arbitrary way.  As a result, the number of clusters based on a point estimate of the cluster configuration using, for example, the {\tt salso} {\tt R}-package (\citealt{salso}) results in 6 clusters for $U \ge 6$. For each model fit we also provide the following $U$-adjusted mean squared error (MSE)
\begin{align}\label{eq:mse}
   \mbox{mse} = \frac{1}{n(K-U)}\sum_{i=1}^n(y_i - \hat{y}_i)^2 
\end{align}
(which compares observed to fitted values taking into account the number of clusters) and the standard deviation of co-clustering probabilities averaged across units.  
\begin{align}\label{eq:sdccp}
  \mbox{sd\_ccp} = \frac{1}{n}\sum_{i=1}^n \mbox{sd}(Pr(z_i = z_{-i} ~|~ \bm{y})).    
\end{align}
This metric measures the cluster ``purity'' as units with co-clustering probabilities that have a larger standard deviation correspond to co-clustering probabilities that are closer to either one or zero compared to those with a smaller standard deviation.  

From Figure \ref{fig:galaxy_ccp} notice that for $U \le 3$ the cluster configuration is quite ``pure'' but with a high mse.  This is not surprising because the galaxy data exhibits three well separated groups, but $K^+=3$ smooths over clear data features resulting in $K^+ >3$ exhibiting a smaller mse.  On the other hand, notice that the nominal number of clusters remains at six even though $mode(K^+ ~|~ \bm{y})$ increases as a function of $U$.  It seems that $U \in \{5,6,7\}$ balances best the quality of cluster configuration and model fit (see Figure \ref{fig:galaxy_ccp}).  

Fitting the aFMM for varying values of $U$ and observing the co-clustering probability matrix for each is  a principled way to study the robustness or uncertainty of the cluster configuration that is easily carried out with the aFMM. 


\subsection{Biomechanic Functional Data Application} \label{sec:biomechanic}
To illustrate the portability of our prior construction among different modeling scenarios, we now employ the aFMM in a functional data example from the field of biomechanics. In this setting, a ``cluster'' is defined to be a collection of curves that are similar in shape and magnitude as defined by a vector of B-spline coefficients.  We briefly introduce the study that produced the data we consider.

\begin{figure}
  \centering
  \includegraphics[scale=0.65]{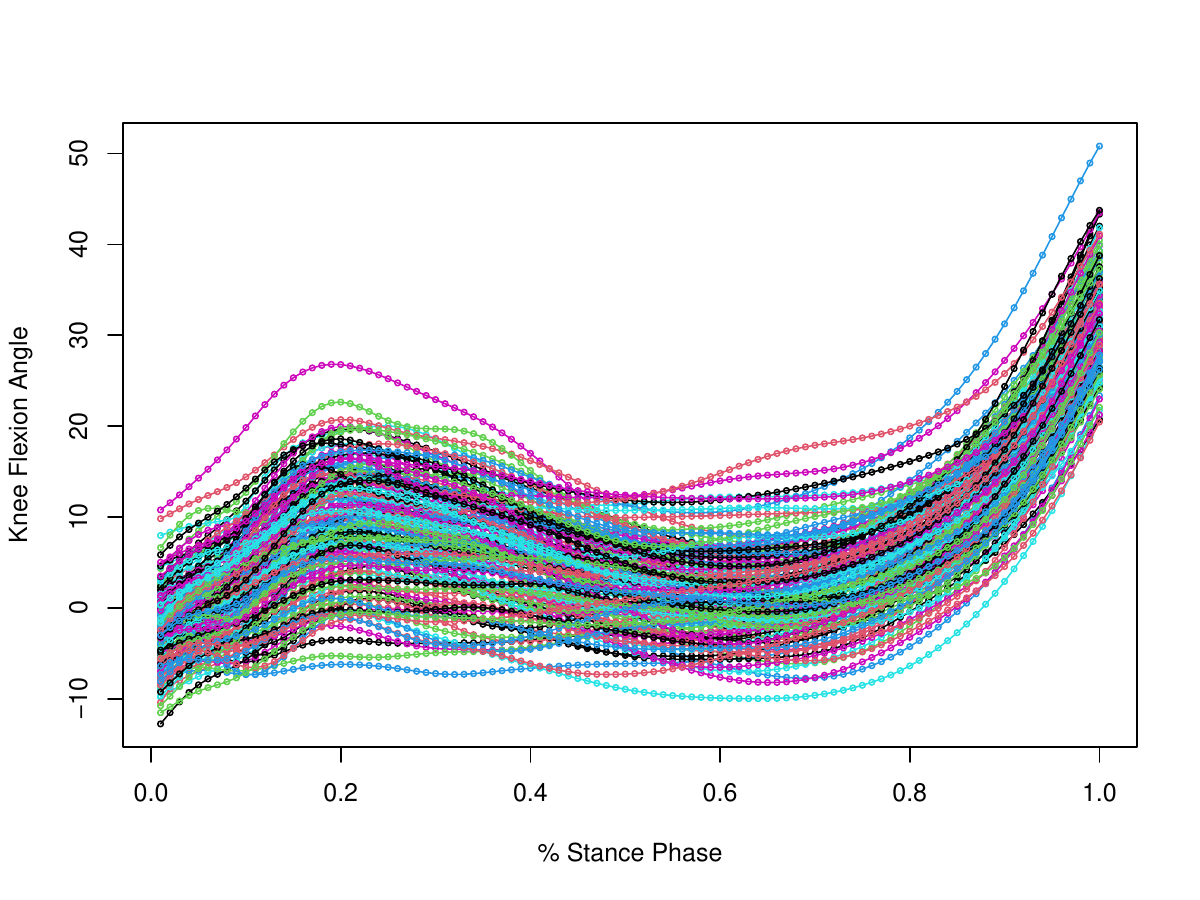}
    \caption{Knee flexion angle for each of the 196 subjects recruited into the study.}
   \label{fig:kfa}
\end{figure}

Biomechanics is the study of how mechanical principles (force and angle) are applied to living organisms.  There is keen interest in learning in what way human biomechanics are connected to joint health.  To this end, 196 subjects that have had reconstructive anterior cruciate ligament (ACL) surgery were recruited to participate in a study that required them to walk on a treadmill.  While walking the knee angle (among other biomechanic variables) was measured through the entire gait cycle (see Figure \ref{fig:kfa}).  Thus, the knee angle measurements could be thought of as discretized functional realizations.  

We seek to identify a subset of movement strategies that subjects adopt post ACL surgery.  
In this study two perspectives and motives for discovering subpopulations exist. First, from a clinician perspective, it would be very useful if a relatively few number of movement strategies are identified as this would facilitate interpretation and treatment formulation (e.g., rigid knee movement, typical knee movement, and flexible knee movement). However, an exercise scientist would not necessarily be concerned with identifying a small number of ``interpretable'' movement strategies, but rather understand the myriad of ways  that the 196 subjects are able to accommodate the ACL surgery.  So a potentially larger number of subgroups would be of interest.  An appeal of the aFMM  is that it can be employed to lucidly handle both situations  through the specification of $U$.  

\subsubsection{Description of Functional Clustering Model} \label{sec:fdamodel}
We employ a functional clustering model that is similar to that detailed in \cite{page&rodriguez&lee_2020}.  For sake of completeness, we detail it here.
Let $\bm{y}_i = (y_{i1}, \ldots, y_{mi})$ denote the $m$ knee angle measurements for subject $i$.  A reasonable functional data model for the data in Figure \ref{fig:kfa} is
\begin{align*}
y_{i}(t) = \beta_{0i} + f_{i} (t) + \epsilon_{i}(t) \ \mbox{where} \ \epsilon_{i}(t) \stackrel{iid}{\sim} N(0, \sigma^2_{i}),
\end{align*}
where $f_i(\cdot)$ denotes the $i$th subject's knee angle function and $\beta_{0i}$ a vertical shift.  Here we assume constant variance at each time point $t \in [0,1]$.  With the desire to flexibly model each subject's curve, we approximate  $f_{i} (\cdot)$ using B-splines which results in the following subject-specific model
\begin{align*}
    \bm{y}_i \sim N(\beta_{0i}\bm{j} + \bm{B}_i \bm{\beta}_i, \sigma^2_i\bm{I}),
\end{align*}
where $\bm{B}_i$ is a $m\times p$ matrix of B-spline basis created by using evenly spaced interior knots and $\bm{\beta}_i$ a $p$-dimensional vector of B-spline coefficients for the $i$th subject.  Curve clustering is then carried out by modeling $\bm{\beta}_i$ with an aFMM
\begin{align*}
    \bm{\beta}_i & \sim \sum_{k=1}^K w_k N(\bm{\theta}_k, \kappa^2_k\bm{I})\\
   \bm{w} & \sim \mbox{Dirichlet}(\bm{\alpha}_{1,2})\\
   \alpha_1 | \alpha_2, U & \sim PC(U, tp=0.1). 
\end{align*}
Smoothing is introduced by modeling $\bm{\theta}_k$ with a penalized B-spline (\citealt{Eilers1996, BayesianPsplines}) under the PC prior framework (\citealt{simpson_etal:2017}) such that
\begin{align*}
Pr(\bm{\theta}_k) & \propto \exp\{1/\tau^2_k \bm{\theta}'_k \bm{S} \bm{\theta}_k\} \\
\tau_k & \sim Exp(\eta_{\tau}).
\end{align*}
Here $\bm{S}$ is a 2nd order random walk penalty matrix and $\eta_{\tau} = -\log(a_{\tau})/U_{\tau}$ where $a_{\tau}$ and $U_{\tau}$ satisfy $Pr(1/\tau_k > U_{\tau}) = a_{\tau}$ {(the PC prior for the  standard deviation of a Gaussian random effect, e.g. $\tau_k$, is the Exponential distribution)}.  Finally,  to balance borrowing-of-strength among units allocated to the same cluster and subject-specific fits, we employ the following priors on the between-subject and within-subject variance components  
\begin{align*}
\sigma_i & \sim UN(0, A)\\
\kappa_k & \sim UN(0, A_0).
\end{align*}
We set $A = 0.001$ as the measured curves are essentially noiseless and $A_0 = 0.25$ which requires clusters to be composed of similary shaped curves.  For $\tau^2_k$ we set $a_{\tau} = 10^{-2}$ and $U_{\tau} = 3.22$ as suggested by \cite{simpson_etal:2017}.  In order to avoid the challenges inherent in multivariate clustering (\citealt{JMLR:v24:21-1276, ghilotti2023bayesian}), we used a small number of interior knots (seven) in the P-spline formulation.  This resulted in $\bm{\beta}_i$ being $p=10$ dimensional which is small enough to not suffer from the curse of dimensionality (\citealt{ghilotti2023bayesian}).   Since the measured knee angle curves are sufficiently smooth each subject's curve is fit well even with 7 interior knots. Lastly, we set $K=25$ for all mixtures that are fit to these data.

To perform clustering from both the clinician's and exercise scientist's perspective we set $U=3$ and $U=10$ with $tp=0.1$?.  For additional context we also fit a sFMM and a static  FMM with  $\alpha=1/K$.  Each of the models were fit by collecting 1,000 MCMC iterates after discarding the first 50,000 and thinning by 100 (this required 150,000 total MCMC samples for each model).  

The posterior distributions of $K^+$ under all four models turned out to be points masses at specific values.   For $U=3$, $mode(K^+ | \bm{y} ) = 6$ which demonstrates that $U$ is indeed a ``soft'' upper bound that can be exceeded when favored by the data.   For $U=10$, $mode(K^+ | \bm{y}) = 9$, while for sFMM and the static FMM  $mode(K^+|\bm{y})=7$.  To further explore the clustering results under each model, we estimated the cluster configuration based on the MCMC samples using the default settings of the {\tt salso} function (\citealt{salso}).  Interestingly the estimated clustering under the four models were all quite different  as all the pair-wise adjusted Rand index (ARI) (\citealt{hubert&arabie:1985}) values between them were less than 0.5.    To further see differences,  we provide Figure S10 of the supplementary material which displays the co-clustering probabilities under each model that was fit.  Clusters are labeled based on subject order.  That is, subject one is always allocated to cluster one, and cluster two begins with first subject not allocated to subject one's cluster, and cluster three begins with first subject not allocated to cluster one or two, etc. Note that there are a subset of subjects that exhibit uncertainty in their cluster allocation, but for the most part the clustering is estimated with low uncertainty.

To visualize the clustering further we provide Figure  \ref{fig:kfa_fits} and Figures S11 - S14 in the supplementary material.  The left column of Figure \ref{fig:kfa_fits} displays the subject-specific curve fits with color indicating cluster memberhship and  the right column displays the cluster-specific mean curves calculated cross-sectionally using all curves allocated to a particular cluster. Note that the subject-specific fits (solid lines through points) are very reasonable for the majority of subjects.   Notice further that one subject was allocated to a single cluster under each model.  It is clear why this is the case as the subjects curve is quite different from the others.  Key differences that exist between the clusters seem to be the height of the knee angle curve early in the stance phase and also the depth of the valley and the sharpness of the drop towards the middle of the stance phase.  It does appear that the desire by clinician's to have 3 clusters forced some subjects whose curves are quite different to be grouped (see Figure S11 of the supplementary material) highlighting the fact that these data highly favor more than three clusters.  The aFMM for $U=10$ generally speaking produced clusters with curves that are more homogeneous relative to the other models.  Cluster seven in the aFMM $U=10$ model would be quite interesting to exercise scientists as it is generally agreed that a shallow valley in the knee angle curve represents ``poor'' biomechanics.  This represents a gate that does not employ much bend at the knee.  Overall, employing the aFMM provides a principled approach to consider both perspectives and the sFMM and static FMM seem to fall somewhere in between the two aFMM models with regards to clusters that exhibit curve homogeneity.


\begin{figure}
  \centering
  \includegraphics[scale=0.8]{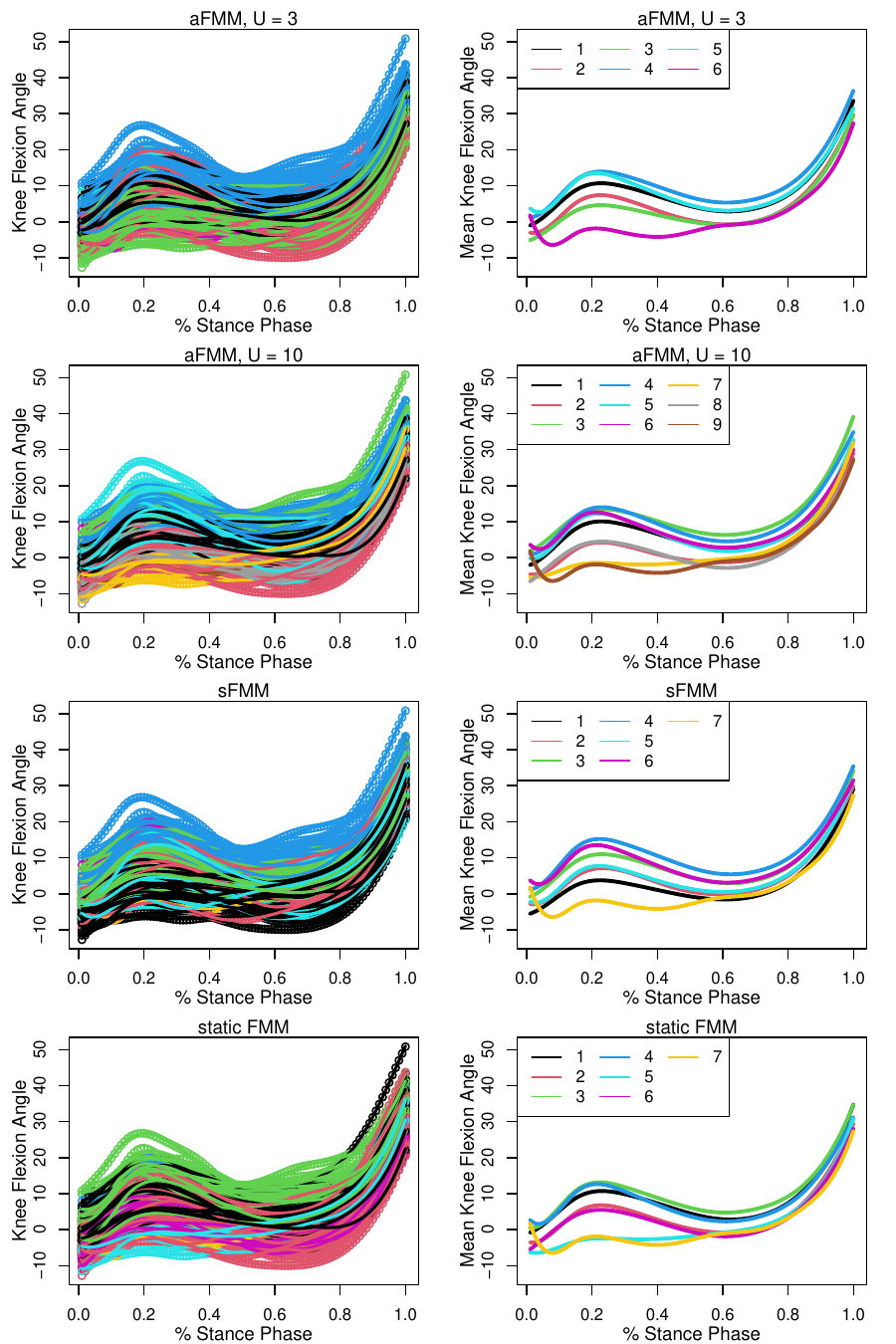}
    \caption{The left column displays the curve fits for all 196 subjects with color indicating cluster (estimated using default settings of the {\tt salso} function).  The right column correspond to the cluster means which were calculated by finding cross-sectional mean among all curves in a cluster.  The first row corresponds to an aFMM with $U=3$, the second an aFMM with $U=10$, the third a sFMM, and the fourth a static FMM.}
   \label{fig:kfa_fits}
\end{figure}

\section{Discussion} \label{sec:discussion}

In this paper we've constructed a prior distribution for arguably the most relevant quantity in model-based clustering; the number of clusters.  This was done by employing an asymmetric Dirichlet distribution as a prior on the weights of a finite mixture.  Further, employing PC prior type technology, we formulated a prior distribution on the shape of the Dirichlet that permits eliciting prior information through intuitive statements that can be asked of the user.   

Our methodology also permits a principled study of the uncertainty associated with the clustering configuration.  The uncertainty associated with $K^+$ can be studied in two ways. The first is through co-clustering probabilities with those that are more ``pure'' indicating a more certain clustering.  Uncertainty can also be explored by studying the stability of the clustering configuration as the value of $K^+$ is changed {\it a priori}.  If either of these two perspectives exhibit uncertainty, then the data are not that informative regarding the number of clusters.  Our prior construction leads to naturally being able to employ both perspectives.

Finally, model-based clustering procedures are, at the end of the day, exploratory approaches that permit users to discover structure in their data.  Our procedure, according to our knowledge, is the first to provide users the ability of carrying out the exploratory data analysis in a principled way based on $K^+$. As a result, the influence that the prior distribution of $K^+$ has on its posterior is something that can easily be studied. 



{\bf Acknowledgements}:  We would like to thank to Christian Hennig for stimulating discussions about the definition of a cluster, Jos\'e J Quinlan for discussions on details associated with the proof of Proposition 1, and the Motion Science Institute at the University of North Carolina for access to the biomechanics dataset.

\singlespacing
\bibliographystyle{asa}
\bibliography{reference}

\end{document}